\documentclass{aa}  

\usepackage{graphicx}
\usepackage{txfonts}
\usepackage{natbib}
\begin{document}

   \title{A reproducible method to determine the meteoroid mass index}

   \author{P. Pokorn\'{y}
          \inst{1}
          \and
          P. G. Brown\inst{1,2}
          }

   \institute{Department of Physics and Astronomy, University of Western Ontario, London, ON, N6A 3K7 Canada\\
              \email{ppokorn2@uwo.ca}
         \and
             Centre for Planetary Science and Exploration, University of Western Ontario, London, Ontario
             }

   \date{Received January 14 2016; accepted April 19 2016}

 
  \abstract
   {The determination of meteoroid mass indices is central to flux measurements and evolutionary studies of meteoroid populations. 
   However, different authors use different approaches to fit observed data, making results difficult to reproduce and the 
   resulting uncertainties difficult to justify. The real, physical, uncertainties are usually an order of magnitude higher 
   than the reported values.}
   {We aim to develop a fully automated method that will measure meteoroid mass indices and associated uncertainty. We validate our method 
   on large radar and optical datasets and compare results to obtain a best estimate of the true meteoroid mass index. }
   {Using MultiNest, a Bayesian inference tool that calculates the evidence and explores the parameter space, we search for 
   the best fit of cumulative number vs. mass distributions in a four-dimensional space of variables ($a,b,X_1,X_2$). We explore 
   biases in meteor echo distributions using optical meteor data as a calibration dataset to establish the systematic offset 
   in measured mass index values.}
   {Our best estimate for the average de-biased mass index for the sporadic meteoroid complex, as measured by radar appropriate to 
   the mass range $10^{-3} > \mathrm{m} > 10^{-5}$ g, was $s=-2.10 \pm 0.08$. Optical data in the $10^{-1} > \mathrm{m} > 10^{-3}$~g range, with the shower meteors removed, produced $s=-2.08 \pm 0.08$.
   We find the mass index used by \citet{Grun_etal_1985} is substantially larger than we measure in the $10^{-4} < m < 10^{-1}$~g range.}
   {}


   \maketitle
%

\section{Introduction}

A fundamental quantity of the dust/meteoroid environment in the solar system is the number distribution of particle sizes with mass. 
The mass or size frequency distribution (hereafter MFD, SFD respectively) of meteoroids provides insight into the origin, evolution, and 
eventual destruction of particles in planetary systems. In particular, the collisional lifetime of meteoroids is intimately linked to 
their present number-mass distribution, and this lifetime is central to dynamical models describing the meteoroid environment \citep{Nesvorny_etal_2011}. 
Moreover, knowledge of the meteoroid mass/size frequency distribution is key to an interpretation of optical and infrared observations of 
the zodiacal cloud \citep{Nesvorny_etal_2010}, the total meteoroid mass influx at the Earth \citep{Plane_2012}, optical and radar meteor flux 
estimates, based on observed rates \citep{Koschack_Rendtel_1990,Kaiser_1960} and empirical models of the interplanetary meteoroid environment \citep{Grun_etal_1985}.

Direct measurement of the SFD/MFD is difficult at small meteoroid sizes. Exoatmospheric detection of meteoroids is currently limited 
to objects of roughly 1 m and larger \citep{Harris_DAbramo_2015}, although meaningful direct estimates of the NEA (Near-Earth asteroid) SFD are not possible below 
a few tens of meters in size owing to the small number of statistics. The SFD/MFD at smaller meteoroid sizes must be inferred from statistical measurements. 

Typically, the MFD/SFD is treated as a free parameter in models and adjusted to fit observations \citep[e.g.][]{Nesvorny_etal_2010}. More 
direct estimates of the MFD may be made through meteor observations in the Earth’s atmosphere, microcrater counting on lunar samples \citep{Morrison_Zinner_1977},
or from in situ measurements \citep{Grun_etal_2001}. 

It is typically assumed that the number of meteoroids, dN, which  have mass between M and M+dM, follows a single mass frequency distribution (MFD),
   \begin{equation}
      \mathrm{d}N \propto M^{-s} \mathrm{d}M\,,
   \label{EQ_1}   
   \end{equation}
where the exponent $s$, the differential mass index is typically between $1.5 - 2.5$ \citep[e.g.][]{Jones_1968}. Integrating Eq. \ref{EQ_1} we
have the total number of meteoroids with mass between $M_1$ and $M_2$ as
   \begin{equation}
      N(M,\alpha) \propto \int_{M_1}^{M_2}{M^{-s}\mathrm{d}M} = \frac{1}{\alpha}
      \left[M_2^{-\alpha}-M_1^{-\alpha}\right] \,,
   \label{EQ_2}   
   \end{equation}
where $\alpha$ is the cumulative mass index distribution exponent and $\alpha = s -1$. Usually, we consider $M_2$ to be orders of magnitude larger than $M_1$; 
theoretically $M_2 \rightarrow \infty$, thus Eq. \ref{EQ_2} could be rewritten as follows:
   \begin{equation}
      N \propto M^{-\alpha} \,,
   \label{EQ_3}   
   \end{equation}
where $N$ is the total cumulative number of meteoroids with mass larger than $M$. The meteoroid population may also be defined by the differential size 
frequency distribution, such that the number of particles, $\mathrm{d}N$, with radius between $r$ and $r+\mathrm{d}r$ is given by
   \begin{equation}
      \mathrm{d}N \propto r^{-u} \mathrm{d}r \,,
   \label{EQ_4}   
   \end{equation}
and where $u$ is the differential size index and $u = 3s-2$ \citep{McDonnell_etal_2001}. For a system in collisional equilibrium, where it is 
assumed all meteoroids have the same strength, $s = 11/6$ \citep{Dohnanyi_1969}, while $s = 2$ represents a meteoroid distribution where mass is equally 
distributed per decade of mass. Physically, the smaller the value of $s$ is, the greater the proportion of large meteoroids in a distribution. Typical values 
for $s$ near the time of maximum of major meteor streams are 1.5--1.9 from both radar \citep{Jones_1968} and visual measurements \citep{McBeath_2015}
at mm sizes. The sporadic background, in contrast, is typically found to be richer in smaller meteoroids than showers, with $s$ ranging from $2 - 2.5$ 
\citep[e.g.][]{Hughes_1978,Thomas_etal_1988,Blaauw_etal_2011} at radar masses (sub-mm sizes to tens of microns). 

Historically, the sporadic mass index was estimated from the distribution of photographic meteor magnitudes. \citet{Hawkins_Upton_1958} used Super-Schmidt 
camera observations to estimate $s=2.34$, a value widely adopted in subsequent literature \citep[e.g.][]{Grun_etal_1985}. However, using a different set of 
Super-Schmidt meteors compiled by \citet{Dohnanyi_1967}, produced an independent estimate of $s \sim 2$, while \citet{Erickson_1968} found $s=2.21$, emphasizing 
the sensitivity of $s$ to both the data used and the analysis methodology. More recently, radar measurements \citep{Baggaley_1999,Galligan_Baggaley_2004} find $s \sim 2$ 
as well, albeit for much smaller masses than for Super-Schmidt data (gram sized vs. microgram). That the mass index changes with particle size is unsurprising, 
but the disparity in measured values for $s$ found in the literature \citep[see][for a good summary]{Blaauw_etal_2011} makes disentangling true variations with 
method- or equipment-specific differences challenging.

Here we present an automated, objective technique for the measurement of meteoroid mass indices, with appropriate uncertainty bounds. We apply this approach to both 
radar data and a suite of previously reduced optical meteor observations with the goal of establishing a best-estimated value for $s$ for the sporadic background 
at $\sim$mg to $\mu$g masses. We extend earlier work by \citet{Blaauw_etal_2011} who also used data from the Canadian Meteor Orbit Radar (CMOR) to estimate $s$ 
by quantifying the role of systematic biases in mass-index measurements from radar data, as first discussed by \citet{Jones_1968}. We empirically test these biases 
directly using known height distributions from optical data, applying equivalent attenuation to the height distributions and by comparing results across multiple 
frequencies, providing both a best global estimate for $s$ and its potential temporal variation. 

\section{Measuring meteoroid mass indices with radar: theoretical considerations}

Meteor radar echoes do not produce direct measurements of an individual meteoroid mass. Rather, some assumptions and model interpretations are required to convert 
the returned radar signal scattered from electrons produced during the ablation phase of meteoroid entry  to some equivalent estimated mass. Here we focus on 
the measurement of mass indices from the meteor amplitude returns recorded by transverse scattering radars.  Radar meteoroid mass distributions may also be estimated 
by measuring the duration distribution of long-lived (overdense) echoes \citep{Baggaley_2002} as well as the returned power from  meteor head echoes \citep{Close_etal_2005}. 

For a backscatter radar, ignoring the effects of fragmentation, it can be shown \citep{McKinley_1951} that the amplitude received from an underdense 
meteor trail in a specular scattering is proportional to the electron line density $q$ averaged over the first Fresnel zone along the trail. This is typically a 
distance of order of a kilometer (or less). Typical meteor trails are an order of magnitude larger than the first Fresnel zone and hence the scattering point may 
fall randomly along the ionization profile, depending on the scattering geometry. In a statistical sense, it has been previously shown \citep{Jones_1968, McIntosh_Simek_1969,Blaauw_etal_2011,Weryk_etal_2013}
that the amplitude distribution can be used as a proxy for the initial meteoroid mass such that the distribution of radar echo amplitudes follows
\begin{equation}
     N \propto A^{-s-1} \,,
   \label{EQ_5}   
\end{equation}
where $A$ is the peak radar amplitude of the echo and $N$ is the cumulative number of echoes with peak amplitude greater than $A$, assuming there is no change 
in $s$ across the dynamic range of masses encompassed by the equivalent amplitude range of the radar. Hence, by recording the amplitude distribution from a shower 
or for the sporadic background, the slope of a plot of log $N$ vs. log $A$ will simply be $1-s$.

Complicating this simple picture are biases inherent to meteor radar echo detection. Though several effects tend to reduce both the echo rate and individual 
echo amplitudes \citep[see][for an excellent summary]{Galligan_Baggaley_2004} the most important of these is typically produced by the initial trail radius (ITR). 
This effect, produced when scattered radio waves from the front and back of the meteor trail scatter out of phase and reduce the power reflected back to the 
radar \citep{CampbellBrown_Jones_2003}, tends to preferentially hide smaller and faster meteoroids from detection as these ablate at higher altitudes 
\citep{Jones_CampbellBrown_2005}. The net effect is that the measured mass index is  lower than the true mass index. Both \citet{Jones_1968} 
and \citet{McIntosh_Simek_1969}  recognized this bias and attempted to make corrections based on estimates of the initial radius that was then available 
and assuming single body ablation. They found that in some cases the correction from apparent to true mass index could be as much as $+0.5$, a significant 
difference. We  explore this effect on our estimates to quantify the systematic errors using a range of modern estimates for the initial radius. 

\section{The Canadian Meteor Orbit Radar: equipment and analysis procedures}

The majority of the data for this study was gathered between 2011--2015 by the Canadian Meteor Orbit Radar (CMOR). Technical details of CMOR can be 
found elsewhere \citep{Jones_etal_2005, Brown_etal_2008}, but we summarize the main features of the system relevant to this work. 
CMOR is a tri-frequency broad beam (55 degree width to 3 dB points) vertically directed radar, simultaneously operating at 17.45 MHz, 29.85 MHz, and 
38.15 MHz with 6 kW, 12 kW, and 6 kW peak power respectively. All three systems have interferometric capability based on the \citet{Jones_etal_1998} 
crossed-antenna receiver array design. This allows CMOR to constrain the direction to each detected echo, with sub-degree accuracy for each system. 
All three systems operate at 532 pulses per second with 3 km range sampling. Echo detection is performed by taking the incoherent sum of 14 pulses 
and searching for excursions in signal amplitude at each range gate that are more than 8$\sigma$ above the noise background. This typically results in a 
detection every $\sim$4 seconds at 29.85 MHz and every $\sim$6 seconds at 38.15 MHz. Each detection has a start time, peak amplitude, echo interferometric 
direction, height, range, and noise level recorded. Details of the interferometry algorithms used, the basis for uncertainty estimates and the detection 
algorithm are described in \citet{Weryk_Brown_2012}. 

\begin{figure}
   \centering
   \includegraphics[width=\hsize]{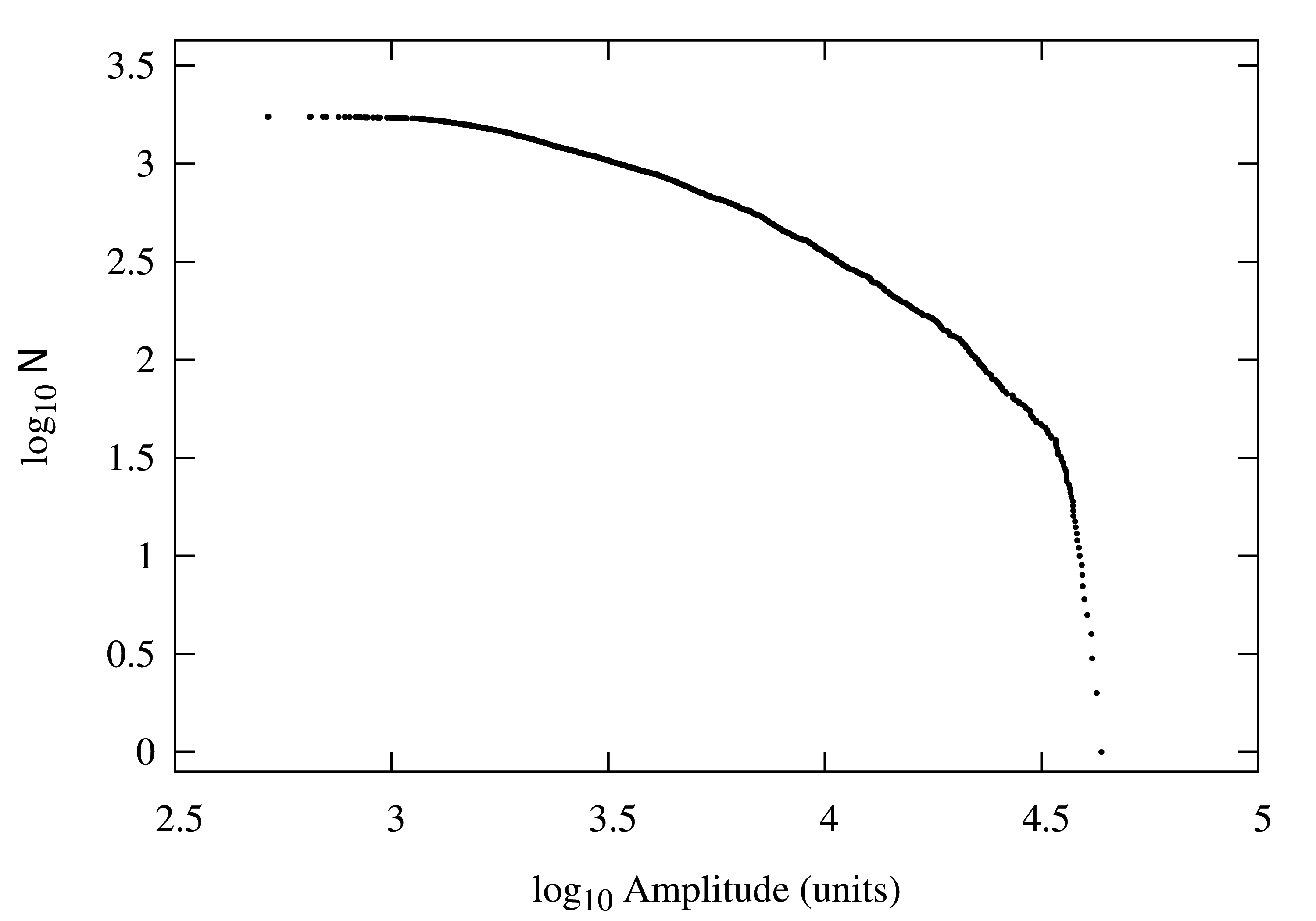}
      \caption{Logarithm of the cumulative number of echoes log $N$ vs. log $A$ for one solar longitude bin ($\lambda=283^\circ$ in 2012) as measured 
      at 29.85 MHz for probable Quadrantid echoes.
              }
         \label{FIG_1}
   \end{figure}

The threshold mass of meteoroids detected depends on speed, but the limiting value at 30 km/s for single-station detection is slightly larger than 
$10^{-8}$ kg corresponding to particles with diameter above 0.3 mm.
We further filter the raw detections by retaining only echoes with heights between 70--120 km. Detections are binned per degree of solar longitude. 
Figure \ref{FIG_1} shows an example of the resulting logarithm of the cumulative number of all echoes log $N$ vs. log $A$ for one solar longitude bin 
as measured at 29.85 MHz. In this figure we also select only meteor echoes that have interferometric directions within $\pm$5 degrees perpendicular 
to the Quadrantid shower. 

At the lowest amplitudes, the cumulative distribution rolls-off to a near constant value as low amplitude events are missed. In practice, this occurs 
near an amplitude of $\sim 10^3$ for CMOR. If all echoes were underdense, the log $N$ vs. log $A$ distribution should produce a straight line of slope 
$1-s$ as $A \propto q \propto m$. However, we see in Fig. \ref{FIG_1} significant curvature to the log $N$ vs. log $A$ line with pronounced steepening 
of the slope at higher amplitudes. This is due to the transition between predominantly underdense echoes (where $A \propto q$) at lower amplitudes and 
overdense echoes (where  $A \propto q^{0.25}$) at higher amplitudes. When the latter regime is reached, the apparent slope changes more closely to $4(1-s)$. 
CMOR’s limiting sensitivity is only slightly more than one decade in mass into the underdense regime, so only the leftmost part of the plot is dominated by 
underdense echoes. The percentage of echoes that  have overdense or transition line densities is further smeared toward the left part of the plot by 
inclusion of a broad set of ranges in the distribution. As the echo range increases, the overdense/underdense limiting amplitude becomes smaller and 
the plot curvature rolls leftward. 

\begin{figure}
   \centering
   \includegraphics[width=\hsize]{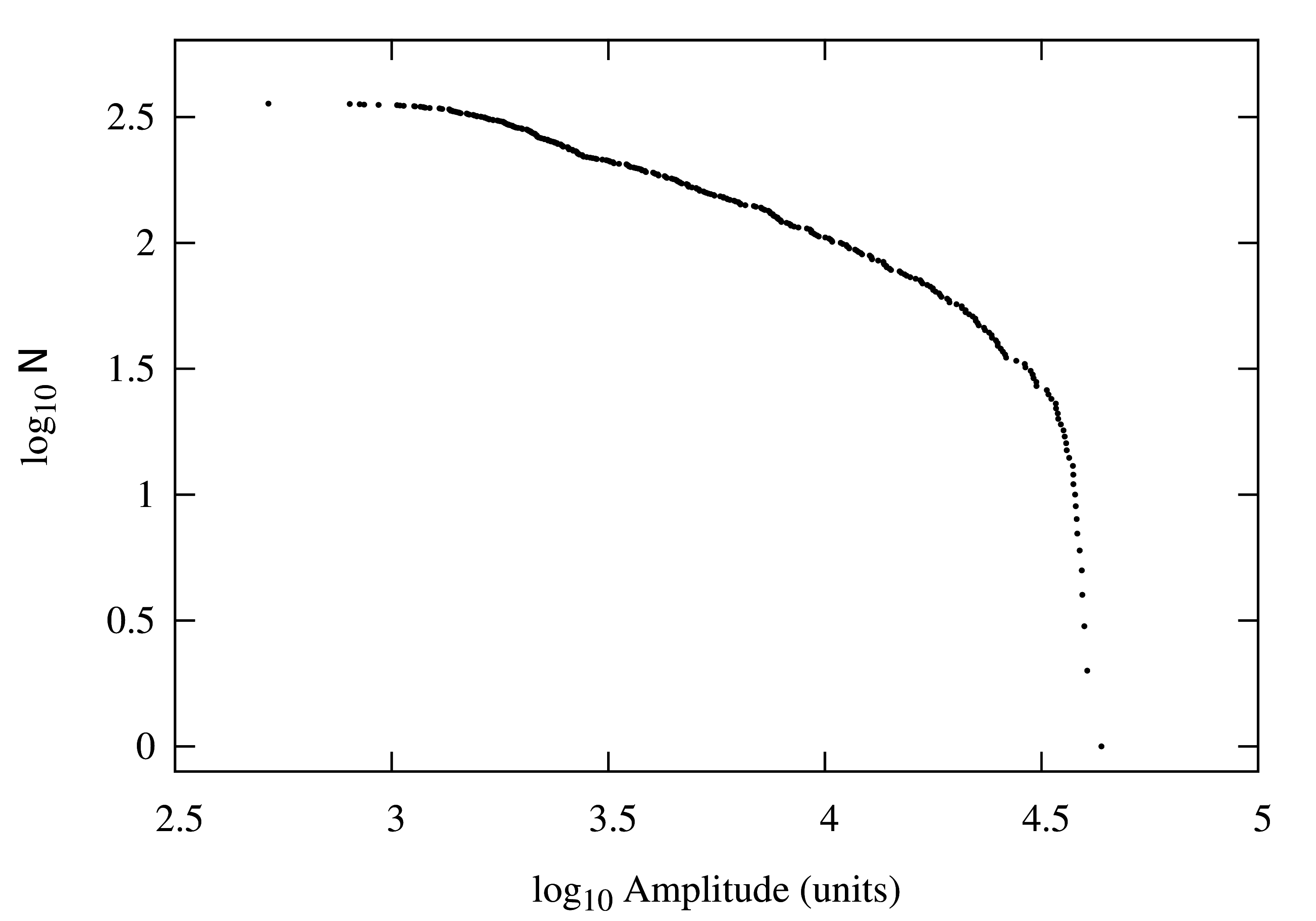}
      \caption{Same echoes as Figure 1, but now with a 110--130 km range filter imposed. 
              }
         \label{FIG_2}
   \end{figure}

\citet{Blaauw_etal_2011} recognized this effect in an earlier CMOR analysis and found that it could be minimized by selecting only echoes with ranges 
between 110--130 km from the radar. This range filter together with the broad CMOR gain pattern tends to freeze the amplitude at which the underdense-overdense 
transition occurs near $\sim 10^4$ amplitude units. Figure \ref{FIG_2} shows the same data as Fig. \ref{FIG_1}, but with the 110--130 km range filter imposed. 
All subsequent number-amplitude analyses use this range filtering.

The flat underdense portion of the plot becomes clearly visible. However, the exact amplitude limits where a fit should be applied and the resulting value 
for $s$ have historically been made through subjective means of a least-squares fit \citep[e.g.][]{McIntosh_Simek_1969, Blaauw_etal_2011}. As this is a 
cumulative plot, the individual data points are correlated and thus the uncertainty found from a least-squares fit is actually a significant underestimate 
of the true uncertainty. The real uncertainty is dominated by the amplitude limits chosen for the fit. In this respect, an objective technique is required to 
fit the flat underdense portion, as well as provide a more physically meaningful estimate of uncertainty. 

\section{Analysis methodology: MultiNest }

For the estimation of uncertainty of our cumulative echo number vs. amplitude fits, we  adopted a software package called MultiNest 
\citep{Feroz_Hobson_2008, Feroz_etal_2009, Feroz_etal_2013}. MultiNest is a Bayesian inference tool with the ability to calculate the evidence, 
with an associated error estimate, and provides posterior samples even for complex multiple-mode distributions.  

To use Multinest, we have to define our problem and provide the likelihood function that will allow the code to compare different sets of solutions. 
Since we assume that the distribution of the cumulative number of echoes above a certain amplitude threshold, $N$, follows a single power-law our 
task is reduced to fitting the first-degree polynomial (line) to a collection of data points in log-log space. The first degree polynomial in our case 
is defined as: $\log_{10}{N}=f(A) =a\log_{10}A + b$, with the slope $a$ (equivalent to $1-s$ if the fit is entirely in the underdense regime) and the 
offset $b$. Fitting the first degree polynomial to a selected dataset is usually a simple task, however, our situation is more complex. Owing to different 
proxies, observing biases/limitations and echo geometries, we do not know where the lower and upper boundaries of our amplitude fitting range are located. 
Thus, we define $x_1$ as the lower amplitude  bound, and $x_2$ as the upper amplitude bound of our fitting range. Our originally two-dimensional problem 
becomes four-dimensional, with one dimension being constrained since $x_1 < x_2$. After several initial tests, we found that logarithmic binning (i.e. 
uniform binning of echo amplitudes in log-log space) of A provides two significant benefits: 1) the fitting procedure runs faster, where the speedup 
depends on the number of meteors in the initial data set, and 2) a natural weighting can be applied to every bin, based on the number of echoes in a given bin. 

To test the goodness of the fit we introduce a modified version of a weighted chi-squared:
   \begin{equation}
      \chi^2 = \left( \frac{N_\mathrm{tot}}{N_\mathrm{bin}}    \right)^{2} N_\mathrm{obs}
      \sum_{n=1}^{N_\mathrm{bin}} \frac{ \left[  \log_{10}N(n) - (a \log_{10} A(n) + b)   \right]^2 } {\sigma^2(n)} \,,
   \label{EQ_6}   
   \end{equation}
where $N_\mathrm{tot}$ is the total number of bins in the dataset, $N_\mathrm{bin}$ is the number of bins within the range ($x_1,x_2$), $N_\mathrm{obs}$ is the 
number of observed meteor echoes, $n$ is the index of the particular bin within the range ($x_1,x_2$),  $A(n)$ is the peak amplitude of the bin, and 
$\sigma^2$ is the variance equal to the number of events in a particular bin. We added the first coefficient in Eq. \ref{EQ_6} to prioritize fitting 
over the longest range in ($x_1,x_2$) possible, while maintaining considerable goodness of the fit. Without this 
additional weight, the best solution is usually the shortest interval that is allowed by our fitting settings (e.g. disregarding fits with five or 
less bins). We tested different powers of the ($N_\mathrm{tot}/N_\mathrm{bin}$) term and found the most consistent results were achieved with a 
square term. For CMOR, our 15-bit ADC (analog-to-digital converter) allows amplitude ranges from 1 to 32 768 (i.e. from $10^0$ to $10^{4.515}$). This range is divided into 5 000 
uniformly spaced bins in logarithmic space.

Multinest effectively searches through our four-dimensional space using a set prior distribution and calculates Bayesian evidence with very high accuracy. 
Our tests indicate that even when the prior distribution is very vague the code is still able to converge to the correct solution. For the slope $a$ 
we inspect the range $(-2,0)$, which translates to the differential mass index range $s=(3,1)$ that covers all reported meteor mass indices 
\citep[e.g.][]{Elford_1968}. The selection of the offset $b$ is arbitrary, depending on the number of observed meteor 
echoes in the dataset simply reflecting the number of observed meteor echoes in the dataset.
For CMOR with echo numbers spanning from 100 up to 5 million. we found the usable range to be $b=(1,12)$. For $x_1$ and $x_2$ (i.e. log $A$), we use 
the range $(2,5)$, since there were no echoes with amplitude $A$ lower than 100 since this is well below the noise floor for CMOR. For all four priors, 
we use a uniform distribution within their ranges.

From our tests, we also found the optimal internal settings required to ensure MultiNest would converge to good solutions.  The maximum number of 
live points is set to $1 600$. We find this number is the best trade-off between accuracy of the solution and the speed of the code. The evidence 
tolerance factor is set to $0.01$. We do not limit the maximum number of iterations that the code is allowed to perform. In some more complicated 
cases, or for very bad datasets, the code might be very slow in finding the solution; however, we did not encounter any infinite loops or runs longer 
than a few hours. 

\begin{figure}
   \centering
   \includegraphics[width=\hsize]{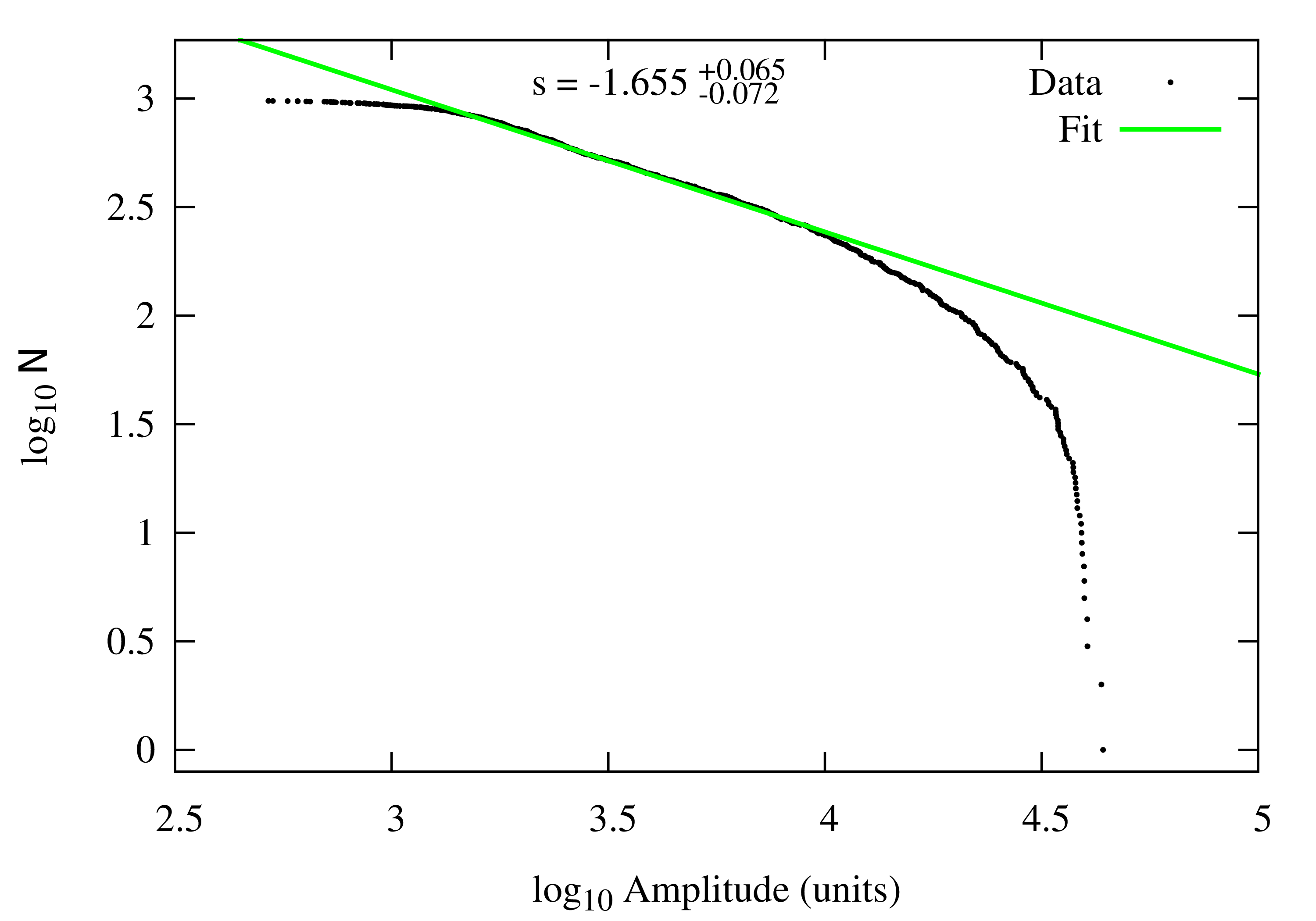}
      \caption{Logarithm of the cumulative number of observed meteors $N$ versus the logarithm of the amplitude $A$ for all Quadrantid 
      radar meteor echoes in 2012 detected during  the three solar longitudes $282^\circ - 284^\circ$ by CMOR. Black dots represent individual 
      echoes and the green line is the best linear fit to the data found by MultiNest. 
              }
         \label{FIG_3}
   \end{figure}

The final result of the search is a set of equally weighted posterior samples that enables us to determine parameters for the best fit, the standard 
deviations of the parameters, and also the local logarithmic evidence of the fit.  Figure \ref{FIG_3} shows results of the Multinest fitting procedure 
applied to the Quadrantid meteor shower in 2012. Here all echoes recorded on 29.85 MHz between $\lambda=282^\circ$ and $\lambda = 284^\circ$ that have 
echo directions within 5 degrees of specular to the Quadrantid radiant were selected. This procedure  includes some sporadic meteors, but given 
the strength of the Quadrantids at maximum, the majority of echoes selected using this specular condition should be from the shower as shown by \citet{Blaauw_etal_2011}.  

\begin{figure*}
   \centering
   \includegraphics[width=17cm]{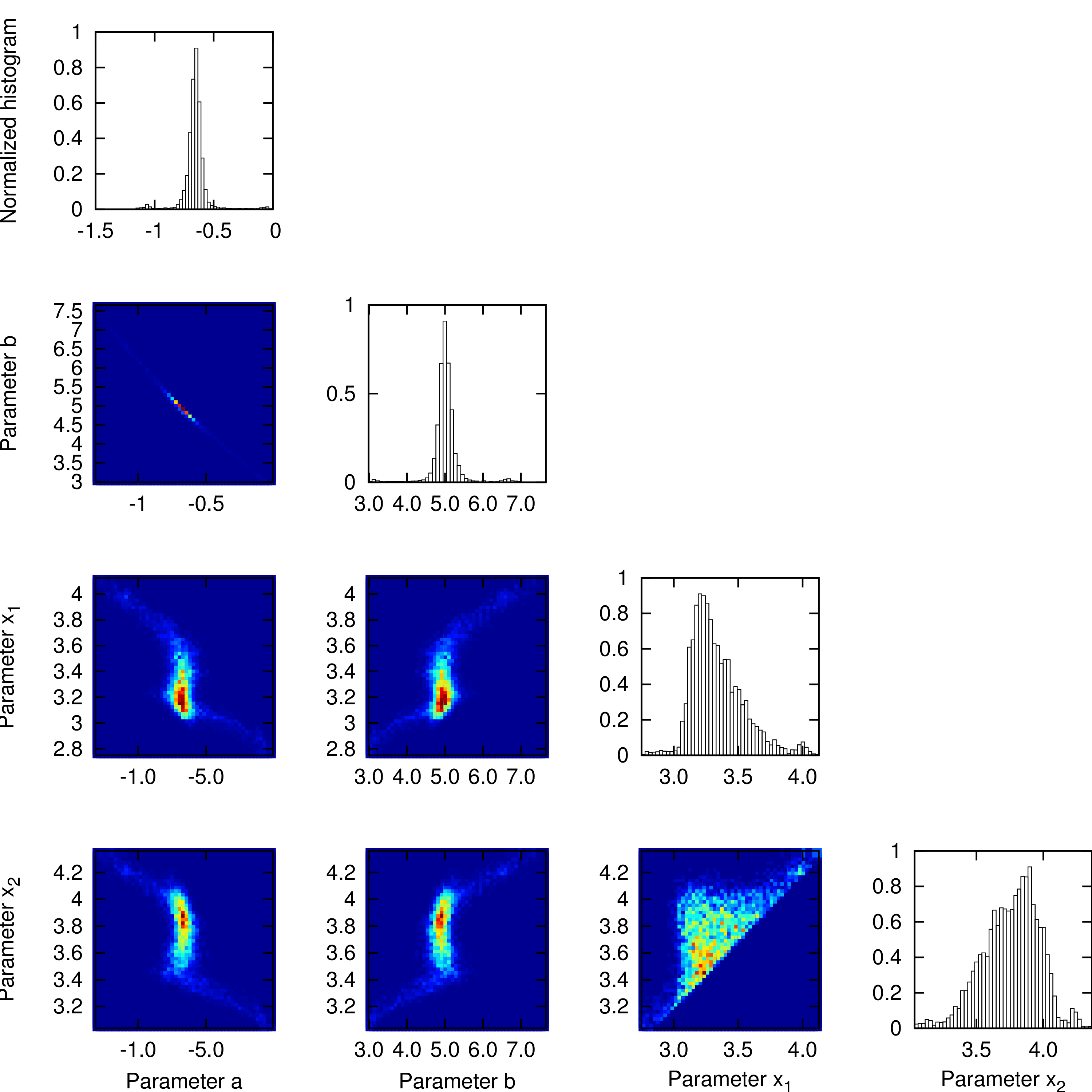}
      \caption{Posterior distribution of four parameters used in our fitting procedure as applied to the Quadrantids meteors shower. Histograms in the outer plots show a distribution of 
      statistically equivalent solutions. The plots inside the triangular structure show projections of these solutions onto two-dimensional 
      planes of two selected parameters, where the color coding  represents the density of solutions (increasing in density from blue to red). 
      The graphs (left to right columns) show the distribution and correlation of the following parameters: (1) slope $a$, (2) offset $b$, (3) lower bound $x_1$, and (4) upper bound $x_2$. 
              }
         \label{FIG_4}
   \end{figure*}

Figure \ref{FIG_4} shows the distribution of weighted posterior samples for all parameters, together with correlations among all parameters for the 
data in Fig. \ref{FIG_3}. The resulting fit in this case is well determined with a very low slope uncertainty, a fact visually apparent from 
Fig. \ref{FIG_4}, since the curve exhibits a very long linear portion without significant rounding. Additionally, all parameters are tightly correlated, an indication of a very good fit. 

\section{Results and discussion}
\subsection{One year of data from 29.85 MHz and 38.15 MHz }

We apply our method to investigate the annual variation of the mass index $s$ of the sporadic meteoroid background during 2014, as measured 
independently by the 29.85 MHz and 38.15 MHz CMOR radars. In our analysis we use all meteors recorded by the main station without any 
constraints for the radiant position or incident velocity, an approach which provides very high statistics (of order $10^4$ or more total 
echoes each day, but only 20\% of these fall into our 110--130 km range bin) but is not suitable for isolating activity from meteor showers 
or specific sporadic meteor sources. Rather, these measurements represent the overall mass index of the entire meteoroid population on average 
to the limiting sensitivity of CMOR. 

\begin{figure*}
   \centering
    \includegraphics[width=17cm]{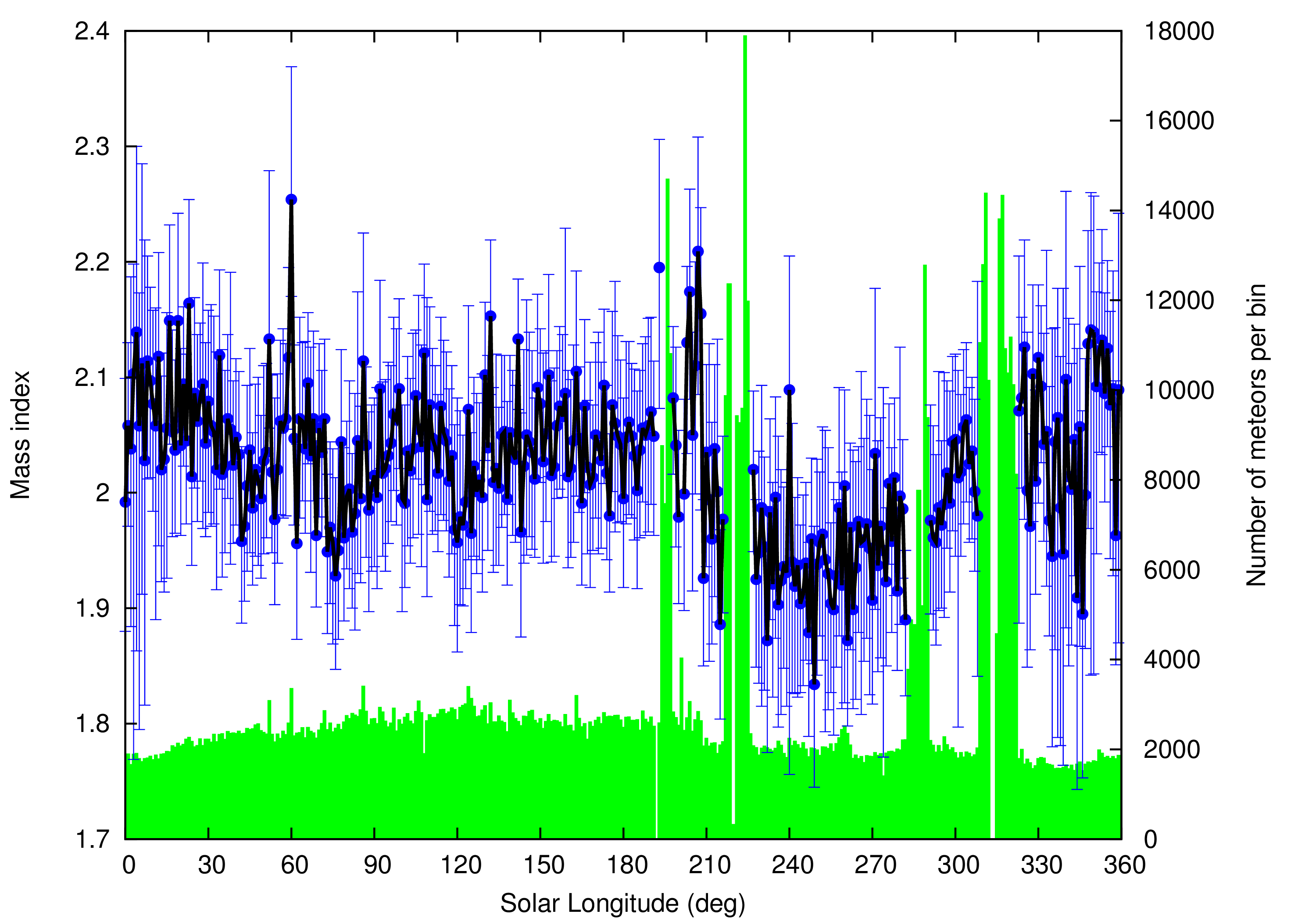}
      \caption{Mass index distribution throughout the whole year 2014 for the sporadic meteoroid complex observed by the 29.85 MHz CMOR radar. 
      Blue points represent the mass index $s$ for a given one degree bin in the solar longitude, where error bars are obtained from the Bayesian 
      posterior equal weight distribution. Green boxes represent the number of meteors in each of one degree bins in the solar longitude. The intervals 
      with large green spikes represent interference and are removed from the analysis.
              }
         \label{FIG_5}
   \end{figure*}

In Fig. \ref{FIG_5} we show the measured mass index $s$ (black solid line in 
Fig. \ref{FIG_5}) for every degree of the solar longitude $\lambda$
in 2014  with its associated uncertainty (blue error bars in Fig. \ref{FIG_5}) based on our analysis, which is applied to data collected by 
the 29.85 MHz radar. We also show the number of observed meteors (in range bins between 110--130 km) for every degree of $\lambda$ (green bars 
with the corresponding y-axis on the right side in Fig. \ref{FIG_5}) during 2014. Several days showed local interference, which dramatically 
increased the number of raw observed meteors (green spikes in Fig. \ref{FIG_5}); we excluded such days from our analysis (gaps in the black 
solid line in Fig. \ref{FIG_5}). 

We find that $s$ remains almost constant during the whole year, except for a noticeable dip at $\lambda =250^\circ$. Averaging these individual 
values for the whole year produces $s=-2.024 \pm 0.064$, where the error is the standard deviation of measured values. If we exclude the region 
around the observed dip $(\lambda=230^\circ-270^\circ),$ we obtain a slightly different average value $s=-2.036 \pm 0.056$. Typical individual 
solar longitude uncertainties of $s$ throughout the year are between 0.07--0.15.

\begin{figure*}
   \centering
 \includegraphics[width=17cm]{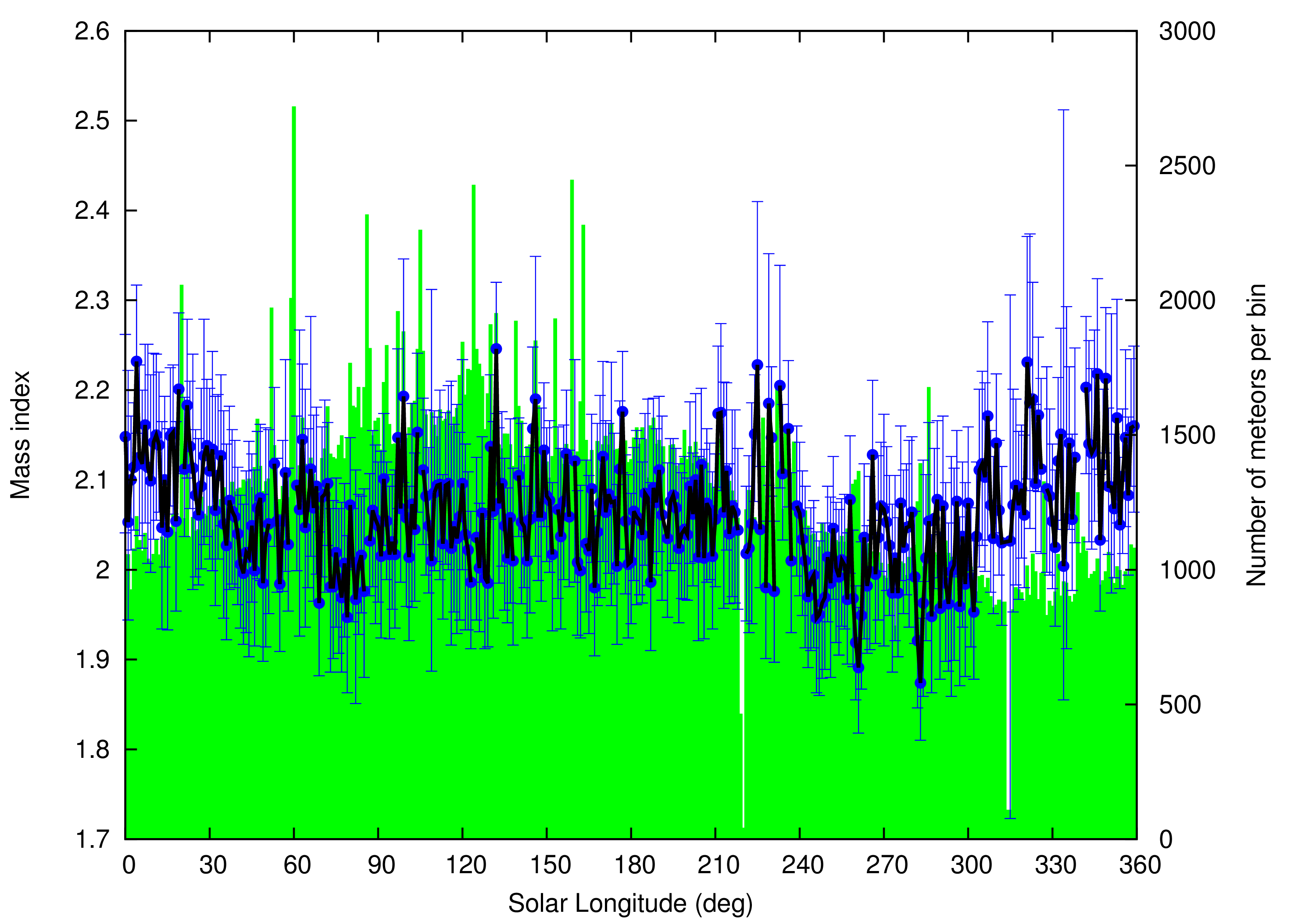}
      \caption{Same as Fig. \ref{FIG_5}, but now for 38.15 MHz single station observations. From the general behavior we see that 29.85 MHz and 
      38.15 MHz single station results are comparable and agree well within determined uncertainty ranges. We note that the mass index for both 
      frequencies has almost identical temporal variations, which agrees with variations of the sporadic meteoroid background previously reported by \citet{Blaauw_2010}.
              }
         \label{FIG_6}
   \end{figure*}

As an independent check of these results, we performed the same analysis for 2014 on data gathered by the 38.15 MHz radar (Fig. \ref{FIG_6}). 
The overall behavior of the mass index during the year is very similar to the 29.85 MHz system with most values agreeing within uncertainty 
(blue error bars in Fig. \ref{FIG_6}).  One difference, however, is that the dip observed or the 29.85 MHz radar is less noticeable and shifted 
by approximately $20^\circ$ to $\lambda=270^\circ$. The average value for the whole year is $s=-2.063 \pm 0.063$; while excluding values of the region 
around the observed dip ($\lambda =250^\circ-290^\circ$) produces $s=-2.071 \pm 0.060$. Individual uncertainties in the value of $s$ 
throughout the year range between 0.8 - 0.15. These values agree within uncertainty with the values found for 29.85 MHz.

As a final check on these average annual values that were computed as a mean of individual daily mass indices, we stacked all data measured during 2014 by 
the 29.85 MHz radar, which produced a single distribution with more than 750 000 measured echoes. The resulting mass index $s$ for the 
stack data is $s=-2.033 \pm 0.08$ identical within uncertainties of our daily average for 2014. 
\subsection{Multi-year data comparison}
We expect the mass index to show similar values and variations year-to-year. This expectation is based on the stationarity of the overall 
sporadic background which, on dynamical grounds, we do not expect to show changes in timescales as short as a few years. However, other 
effects may have influence our measured mass index values, including instrumental effects and long-term changes in the upper atmosphere (particularly 
its mass density). This can affect the ablation behavior of meteoroids (notably their trail length) and hence their radar detectability, an effect 
long recognized as a bias in meteor radar data \citep{Ellyett_1977}. Indeed, such long-term atmospheric density changes have been inferred from 
CMOR measurements \citep{Stober_etal_2014}. 

However, \citet{Stober_etal_2014} and a similar study from the southern hemisphere by \citet{Lima_etal_2015} suggest 
that, during the whole solar cycle, the neutral air density varies by only a few percent, resulting in an annual 
mean meteor peak height as measured by radar changing by no more than one kilometer. Between 2011--2015, solar 
activity as measured by the 10.7 cm flux increased by only 20\% with the most pronounced increase towards the 
end of 2015 \citep{Stober_etal_2014}. It is clear that changes in the mass density of the upper atmosphere 
affect meteor echo counts \citep{Lindblad_1968,Hughes_1976,Ellyett_Kennewell_1980,Elford_1980,Lindblad_2003} 
and likely also affect the begin and end heights of meteor ablation \citep{PellinenWannberg_etal_2010}, but the solar 
cycle impact on mass index measurements by radar is yet to be determined. 

The current CMOR detection and radar-processing pipeline for the 29.85 MHz has been stable since 2011, hence we expand our analysis to 2011--2015 
data to confirm if the same general intrannual trends repeat as expected, if the signal is mainly reflecting intrinsic changes in the mass index.

\begin{figure*}
   \centering
    \includegraphics[width=17cm]{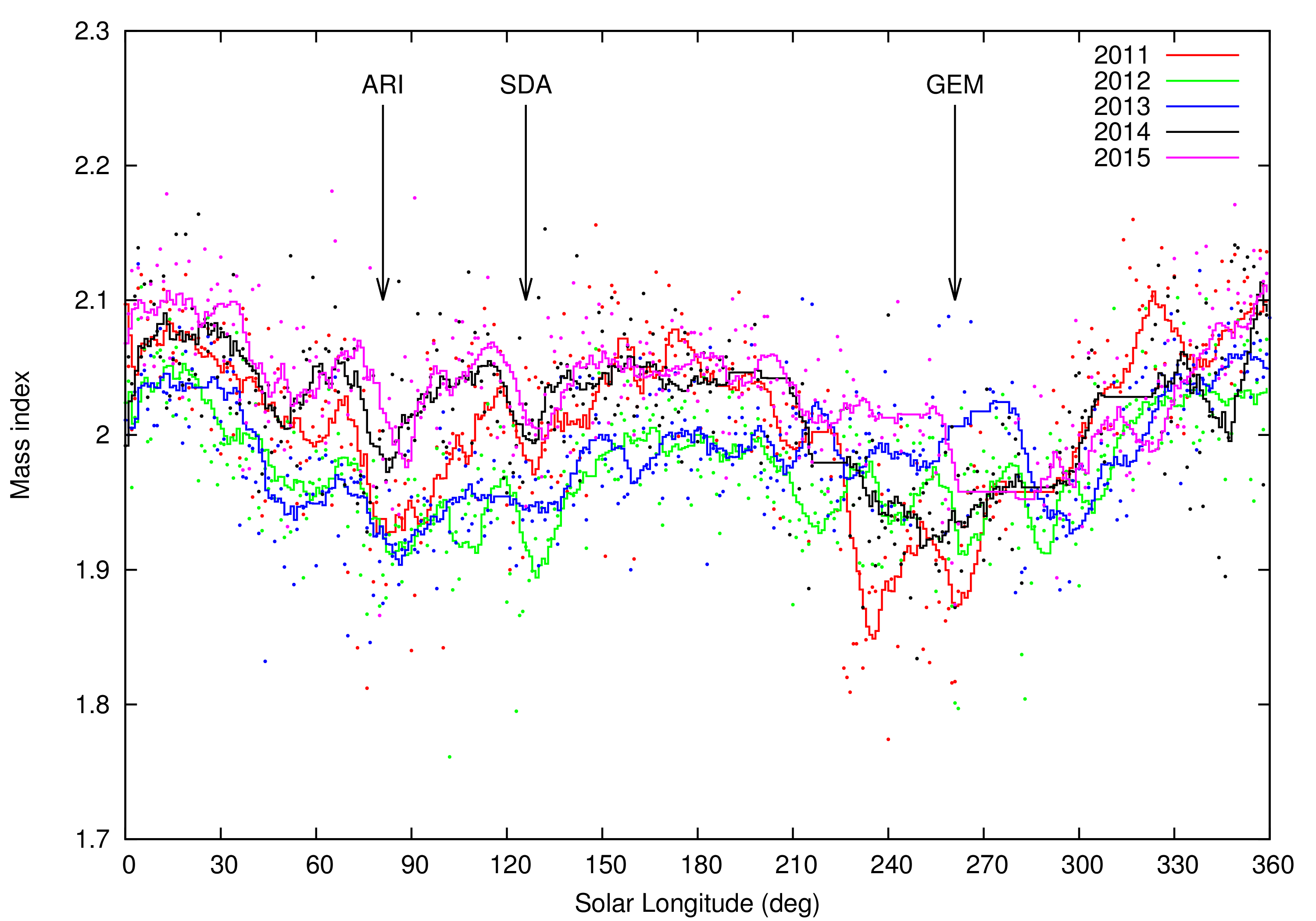}
      \caption{Variations of the mass index measured over five consecutive years 2011--2015 by the 29.85 MHz CMOR radar. Color-coded dots 
      represent the mass index for each $1^\circ$ solar longitude bin and color-coded lines show the moving average with $10^\circ$ window 
      averages. Arrows with labels denote maxima of Arietids (ARI), Southern Delta Aquariids (SDA), and Geminids (GEM) meteor showers.
              }
         \label{FIG_601}
   \end{figure*}

Figure \ref{FIG_601} shows the mass index $s$ variation with the solar longitude (dots color-coded by years) for years 2011--2015, as measured by the 
29.85 MHz radar. To increase the readability of the plot, we omitted error bars for each recorded $s,$ noting that uncertainties remain very similar 
to the year 2014. We introduce a moving average with a $10^\circ$ sliding window (solid lines color-coded by years) to better follow the data trend. 
Examination of Fig. \ref{FIG_601} leads us to conclude: a) the mass index $s$ retains its global features during 2011--2015, though experiencing some 
fluctuations in absolute value, b) the dips observed in 2014 (see Fig. \ref{FIG_5}) are not artifacts and appear during previous years as well. 
We suggest the intrannual fluctuations are due to strong showers in the data, while the absolute interannual changes in the apparent mass index are likely caused by solar 
activity affecting the mass density of the atmosphere at meteor ablation heights. 

\begin{figure}
   \centering
   \includegraphics[width=\hsize]{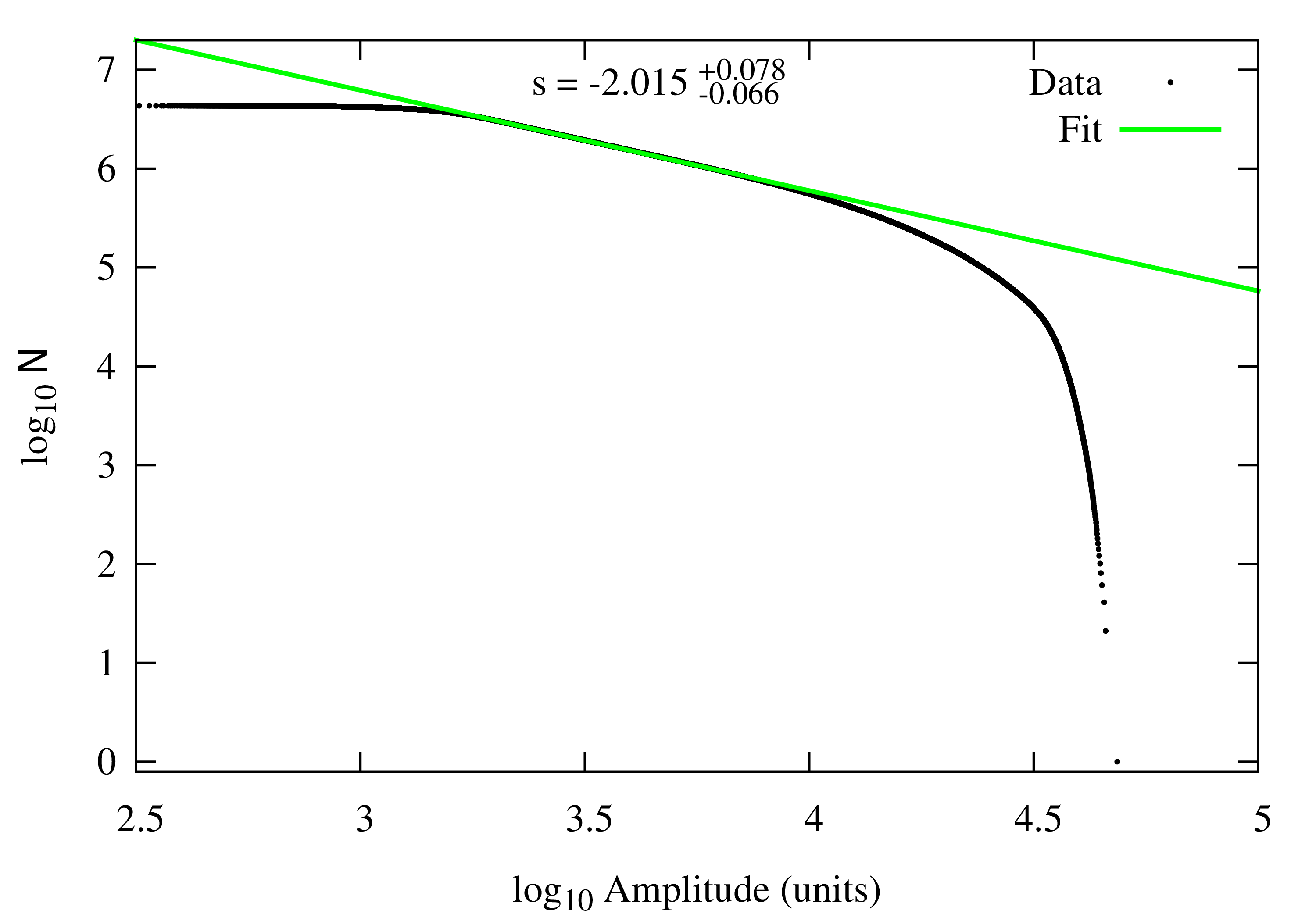}
      \caption{The logarithm of the cumulative number of observed meteors $N$ versus the logarithm of the amplitude $A$ for all
      radar meteor echoes measured over five consecutive years 2011--2015 by the 29.85 MHz CMOR radar. Black dots represent individual 
      echoes and the green line is the best linear fit to the data found by MultiNest. 
              }
         \label{FIG_Althea}
   \end{figure}

From Fig. \ref{FIG_601}, we see that some dips reappear each year at exactly the same time; these dips originate from the strongest annual meteor 
showers, which produce a large number of all detected echoes at the time of their maximum. Using the results of \citet{Brown_etal_2010}, we 
readily associate the dip at $\lambda=45^\circ$ with the Eta Aquariids, $\lambda=81^\circ$ with the Daytime Arietids (ARI), the dip at 
$\lambda=126^\circ$ with the  South Delta Aquariids (SDA), and the final dip at $\lambda=261^\circ$ with the Geminids (GEM). Indeed, these four 
showers were found by \citet{Brown_etal_2010} to be among the top five strongest detected by CMOR. Table 1 summarizes the overall average mass 
index for each year, as well as the cumulative index for all data. Our overall best fit for all 29.85 MHz data is $s=-2.015 \pm 0.072$ (Fig. \ref{FIG_Althea}). 

\begin{figure*}
   \centering
   \includegraphics[width=17cm]{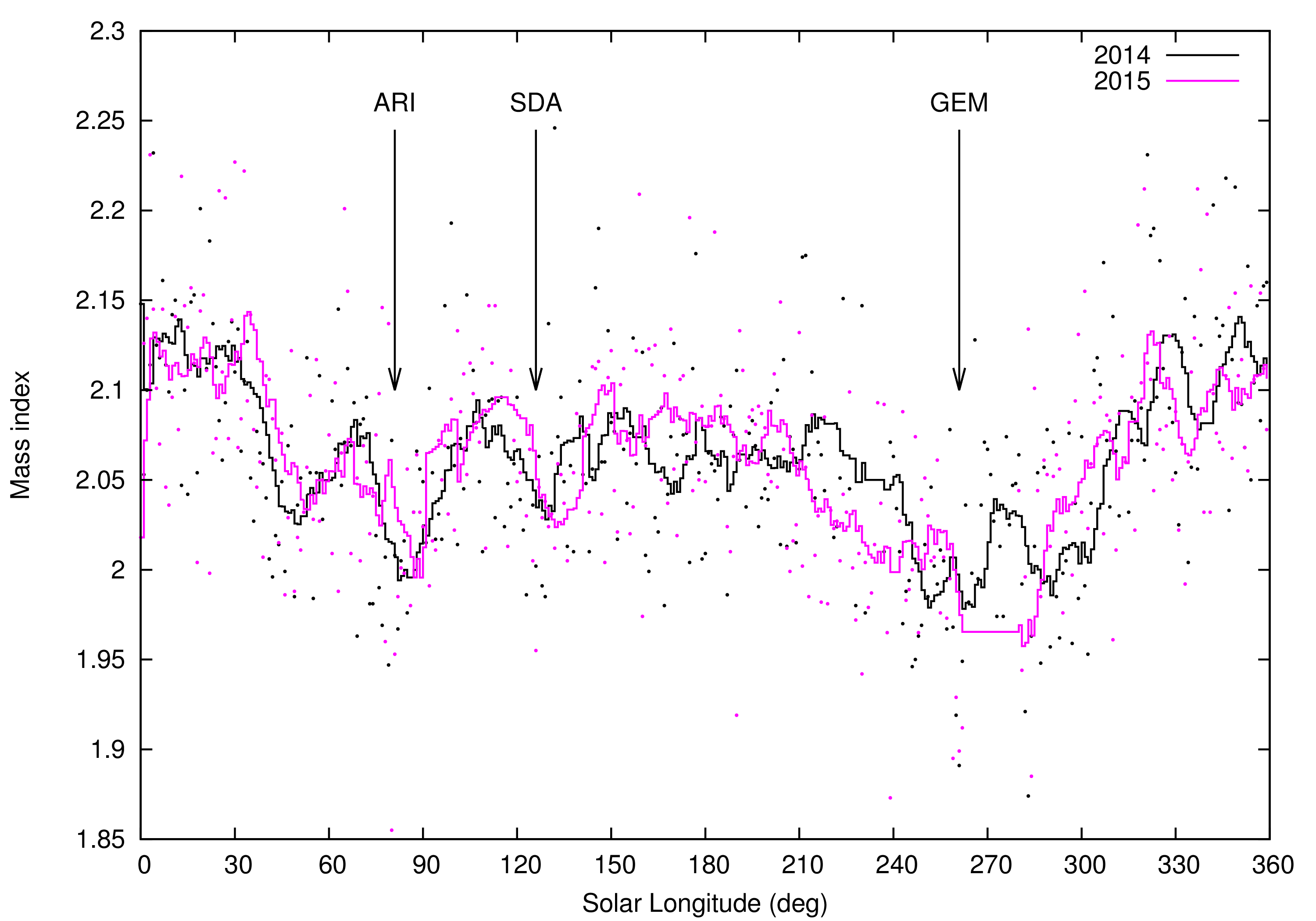}
      \caption{Same as Fig. \ref{FIG_601}, but now for the years 2014--2015 for 38.15 MHz. Years 2011--2013 are unavailable owing to a 
      change in the software detection, which resulted in many non-specular overdense echoes being selected, biasing the mass index value 
      to lower values.}
         \label{FIG_602}
   \end{figure*}

Data measured at 38.15 MHz are shown in Fig. \ref{FIG_602} for the years 2014--2015. The raw data-processing pipeline for 38.15 MHz included many 
non-specular trails prior to Nov 2013 so we exclude earlier data for comparison with 29.85 MHz. The Daytime Arietids at $\lambda=81^\circ$ 
are less noticeable than on 29.85 MHz, while the South Delta Aquariids at $\lambda=126^\circ$ result in a significant drop in $s$. The 2014 
and 2015 average values from Table \ref{TAB_1} for 38.15 MHz are similar to the overall average of $s=-2.080 \pm 0.075$. 

We find 29.85 and 38.15 MHz produce similar mass indices, both as averages and as yearly distribution aggregates. This provides confidence 
that we are not heavily affected by frequency dependent biases in detection. Taken together, the raw radar best estimate for the global meteoroid 
background from CMOR multi-frequency data is $s=-2.05 \pm 0.08, $ which encompasses the spread in the means and the temporal variability.

\begin{table*}
\caption{The measured mass index for each year as a single aggregate distribution for 29 and 38 MHz. Shown for comparison is the mass index 
found for the CAMO influx system (true), together with the apparent change to the mass index when the initial trail radius bias is applied with 
no velocity correction, and with a velocity correction.}             
\label{TAB_1}      
\centering                          
\begin{tabular}{c c c}        
\hline\hline                 
Year & $s$ & $N_\mathrm{echoes}$ \\    
\hline                        
\multicolumn{3}{c}{29.85 MHz} \\
\hline
2011 & $ -2.019 \pm 0.071 $ & 838 648 \\
2012 & $ -1.981 \pm 0.074 $ & 1 032 151 \\
2013 & $ -1.996 \pm 0.074 $ & 942 622 \\
2014 & $ -2.033 \pm 0.082 $ & 749 059 \\
2015 & $ -2.044 \pm 0.075 $ & 758 040 \\
Overall & $ -2.015 \pm 0.072 $ & 4 320 520 \\
\hline
\multicolumn{3}{c}{38.15 MHz} \\
\hline
2014 & $ -2.080 \pm 0.072 $ & 417 381 \\
2015 & $ -2.078 \pm 0.076 $ & 370 658 \\
Overall & $ -2.080 \pm 0.075 $ & 788 039 \\
\hline
\multicolumn{3}{c}{CAMO} \\
\hline
TRUE & $ -2.082 \pm 0.056 $ & 3 106 \\
ITR attenuation with no velocity correction & $ -2.044 \pm 0.036 $ & 3 106 \\
ITR & $ -2.069 \pm 0.035 $ & 3 106 \\\hline                                   
\end{tabular}
\end{table*}

\subsection{Bias owing to initial trail radius effect on mass index measurement}

Our previous result is the apparent mass index of the observed echo population. However, it is well known that the height distribution 
and flux measurements of underdense meteor echoes are biased owing to the initial trail radius (or echo height ceiling) effect \citep{Jones_CampbellBrown_2005}. 
This effect results in severe attenuation of the reflected radio signal when the meteor trail dimension is comparable to the radar 
wavelength. In most cases, echoes with large radii relative to the radar wavelength go entirely undetected. One aspect of this effect is 
that even echoes that are detected will have their amplitudes damped relative to trails of the same electron line density with no initial 
radius and this may affect the apparent mass index \citep{Jones_1968}. The finite-velocity effect \citep{Ceplecha_etal_1998} has a similar 
attenuation on forming trails, but is generally smaller in magnitude than the initial radius effect for CMOR and is ignored here \citep{CampbellBrown_Jones_2003}. 

Previous work has suggested theoretical corrections for apparent mass indices based on initial radius attenuation \citep{Jones_1968,McIntosh_Simek_1969} 
but were based on single-body (non-fragmenting) assumptions about ablation. The results of these studies suggested that the apparent mass index 
would be an underestimate of the true mass index (possibly by as much as 0.5), primarily due to initial radius effects. However, actual meteoroid 
ablation is affected by fragmentation \citep{Ceplecha_etal_1998} and the resulting influence on the mass index is less clear.

As a means for checking our mass determination technique and accuracy of our results without recourse to any model assumptions, we choose an 
empirical approach. We make use of optical meteor observations obtained by the Canadian Automated Meteor Observatory (CAMO) wide-field influx 
camera system \citep{Weryk_etal_2013}, an automated two-station video meteor system located in  southern Ontario that has been operating since 2009. The 
limiting meteor magnitude of the CAMO influx system is +7$^\mathrm{M}$ \citep{CampbellBrown_2015}, which is similar to CMOR, based on simultaneous optical-radar 
meteor measurements \citep{Weryk_Brown_2013}. 

Our intention is to compare the mass and height distribution of all meteors measured by CAMO to obtain the population of meteors that CMOR should 
be able to observe if it were not affected by the initial trail radius effect, i.e. we take the CAMO influx height distribution to represent the 
true distribution that CMOR samples. Using the CAMO-observed height distribution, we apply the initial trail radius attenuation based on 
\citet{Jones_CampbellBrown_2005} to the observed population of CAMO optical meteors and compare the results with radar meteors recorded by CMOR. 
Here we begin with the de-biased data set from the influx cameras as described in \citet{CampbellBrown_2015} of 3 437 manually measured meteors 
with masses in range $10^{-3}$ kg to $10^{-7}$ kg based on a constant luminous efficiency value of 0.7\% \citep{CampbellBrown_Koschny_2004}. In addition, we 
 removed all shower meteors that match any of the established IAU (International Astronomical Union) Meteor showers. This left 3 106 meteors in our optical “sporadic” sample. 

\begin{figure}
   \centering
   \includegraphics[width=\hsize]{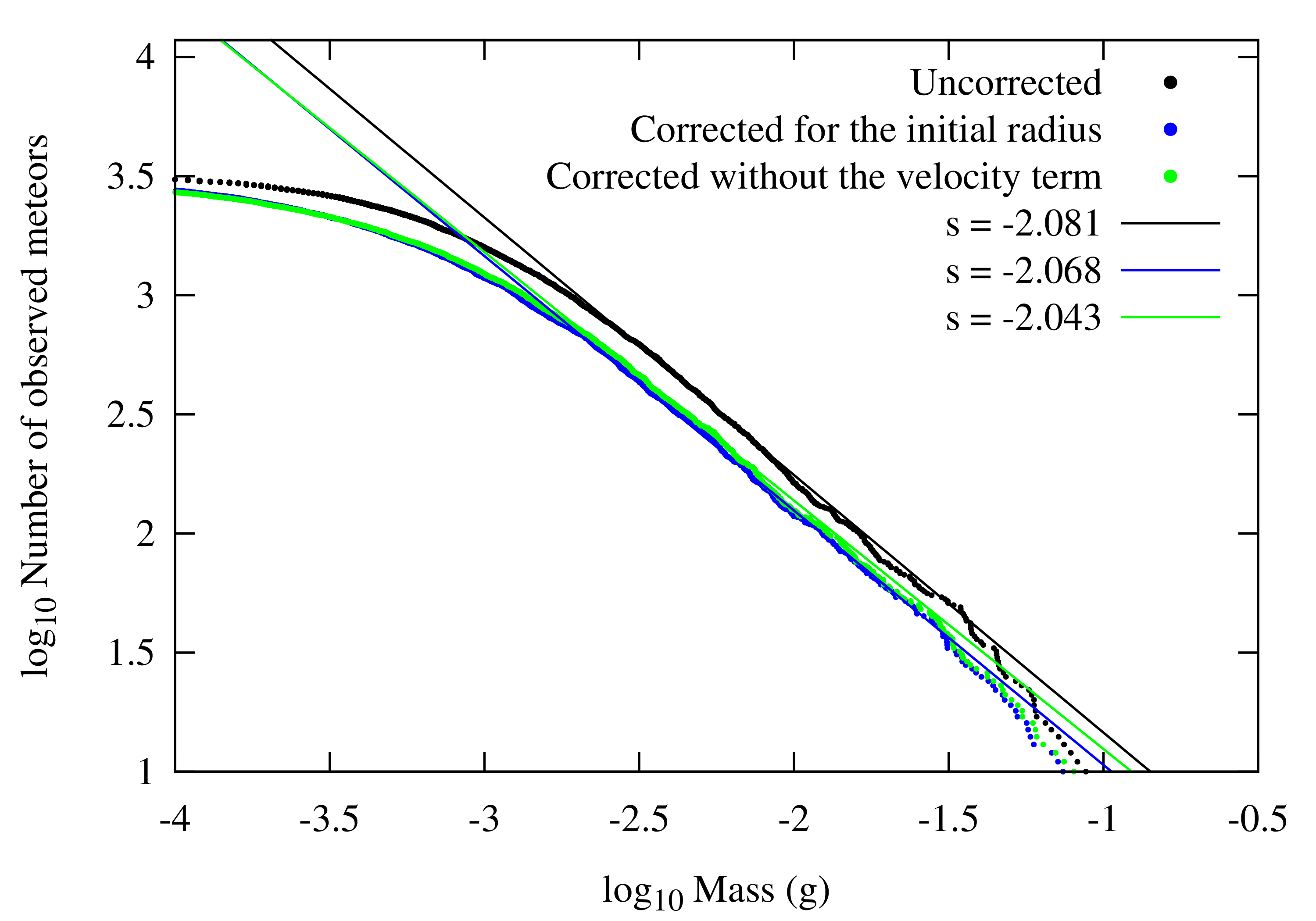}
      \caption{Logarithm of the cumulative number of observed meteors $N$ relative to the logarithm of their mass for meteors observed with 
      CAMO between 2009 and 2012. We show CAMO data not affected by the initial trail radius (black dots) correction with the best fit 
      (black line). This is the true mass index. The distribution once the initial trail radius bias (blue dots) is applied retains an almost 
      identical slope (blue line). We also show the effect of applying the initial radius bias with a velocity term to the distribution and 
      find it produces a negligible difference (green dots and green for the best fit).
      }
         \label{FIG_603}
   \end{figure}

Figure \ref{FIG_603} shows the logarithm of the cumulative number of these 3 106 observed CAMO meteors versus the logarithm of meteoroid mass 
(black points). Applying our MultiNest fitting procedure to this distribution, we obtain $s = -2.090 \pm 0.045$ (black solid line in Fig. \ref{FIG_603}). 
This value agrees within uncertainty with our results obtained from CMOR. This outcome suggests that the CMOR distribution is minimally affected by 
the height-ceiling effect. To check this assertion, we applied the attenuation to the echo amplitude caused by the initial trail radius to all the 
CAMO meteors following the procedure outlined in \citet{Jones_CampbellBrown_2005}, using the optically measured speed and height of peak luminosity 
of the meteor (blue points in Fig. \ref{FIG_603}). Here we use the photometric mass of the CAMO meteoroids as a proxy for the peak radar echo 
amplitude each meteor would produce in the absence of any attenuation effects. Taking the original CAMO distribution and applying the initial 
radius attenuation produces  a small offset in the slope of the mass distribution,  which modifies the apparent mass index to the slightly smaller 
value of $s = -2.043 \pm 0.025$ (blue solid line in Fig. \ref{FIG_603}). This agrees within the respective uncertainties of the previously obtained 
values from radar measurements.  Finally, we used the measured CAMO peak brightness alone to simulate the initial radius correction, excluding the 
effect of the measured velocity (green points in Fig. \ref{FIG_603}). The resulting change is again negligibly different from the measured radar 
values, with a resulting index $s = -2.053 \pm 0.026$.

Based on our empirical checks, the attenuation owing to the initial trail radius effect only has  a minor influence on our measured mass index, 
since it tends to translate the entire curve rather than changing its slope significantly. This change should cause a systematic shift of no more than 0.05
to the larger values. Applying this shift in $s$ to our best fit of observed radar-determined mass index gives us a so-called corrected value $s = -2.10 \pm 0.08$.

\begin{figure}
   \centering
   \includegraphics[width=\hsize]{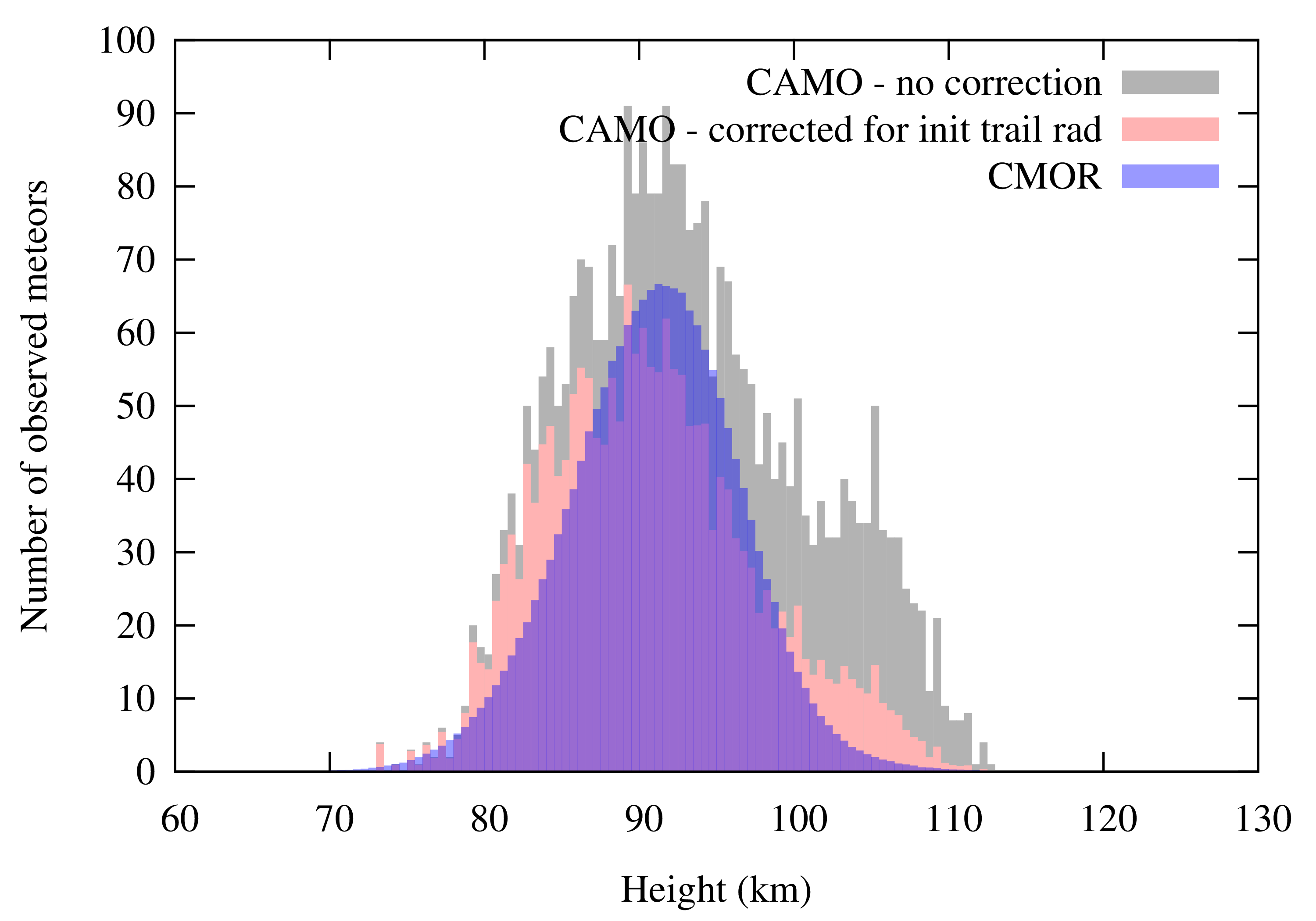}
      \caption{Distribution of heights for uncorrected CAMO data (gray histogram), CAMO data corrected for the initial radius (red histogram), 
      and the 29.85 MHz CMOR heights in 2014 (blue histogram) rescaled to match CAMO corrected distribution. The effect of the correction is 
      evident, minimizing the contribution of meteors with higher heights while keeping meteors with  lower heights untouched.
      }
         \label{FIG_604}
   \end{figure}

However, the initial radius effect is important when considering the observed radar height distribution. 
We can use the CAMO height distribution as ground truth and then apply the attenuation based on peak brightness height and velocity measurements, 
and then compare the resulting height distribution with CMOR measurements. Figure \ref{FIG_604} shows the height distribution of 3 106 CAMO meteors 
(gray histogram), the equivalent effect of attenuation on this population if it were observed by CMOR (red histogram), and finally the observed 
height distribution of 1.5 million meteors measured at 29.85 MHz by CMOR in 2014 (blue histogram). The height distribution of CMOR meteors was 
normalized to match the maximum value of the attenuated population. While the observed CAMO height distribution is very different from observed 
CMOR echo height distribution, the fact that the corrected population is a reasonable match to the CMOR height distribution validates our earlier 
approach of simulating the effects of initial radius on the mass index.  We note that the largest difference between the predicted CAMO-attenuated 
height distribution and the CMOR observed heights between 100--110 km may reflect larger initial radii at these heights than originally found by 
\citet{Jones_CampbellBrown_2005}. This would be consistent with recent optical measurements from \citet{Stokan_CampbellBrown_2014}, who measured much 
larger apparent initial radii at such heights.

\subsection{Mass index comparison with earlier studies}

The most recent study of the meteoroid environment mass index based on CMOR data comes from \citet{Blaauw_etal_2011}. This study uses 
the same radar with 29.85 MHz frequency before it underwent a major upgrade that doubled its power. In principle, their findings should 
be similar to ours, since this upgrade decreased the limiting mass limit by only 40\%. For the whole meteoroid complex during 2007--2010, 
\citet{Blaauw_etal_2011} find  $s = 2.17 \pm 0.07$, slightly higher than our value, but using a manual method for curve-fitting and a 
different meteor-echo detection code. 

Many earlier measurements of the mass index from radar and optical measurements \citep{Simek_McIntosh_1968} found values for $s$ in 
the range $2.2 - 2.5$ in our mass-magnitude range.  The widely used interplanetary meteoroid model flux given by \citet{Grun_etal_1985} 
adopts the mass index $s = 2.34$ for the mass range from $10^{-5}$ to $10^2$g. This is based on the photographic meteor studies by 
\citet{Hawkins_Upton_1958} who found $s = 2.34 \pm 0.06,  $ which was obtained by reducing and fitting a data set consisting of 300 brighter Super-Schmidt 
meteors. This is appropriate to meteors of $-2<\mathrm{M_{Ph}}<+3.5$, corresponding to observed limiting masses of order 0.01--0.1g using the 
mass-magnitude-velocity relation of \citet{Jacchia_Verniani_1967} and a mass weighted mean speed of 17 km/s \citep{Erickson_1968}. This value was 
later used in a comprehensive study of \citet{Whipple_1967} and subsequently widely reproduced.  

\begin{figure}
   \centering
   \includegraphics[width=\hsize]{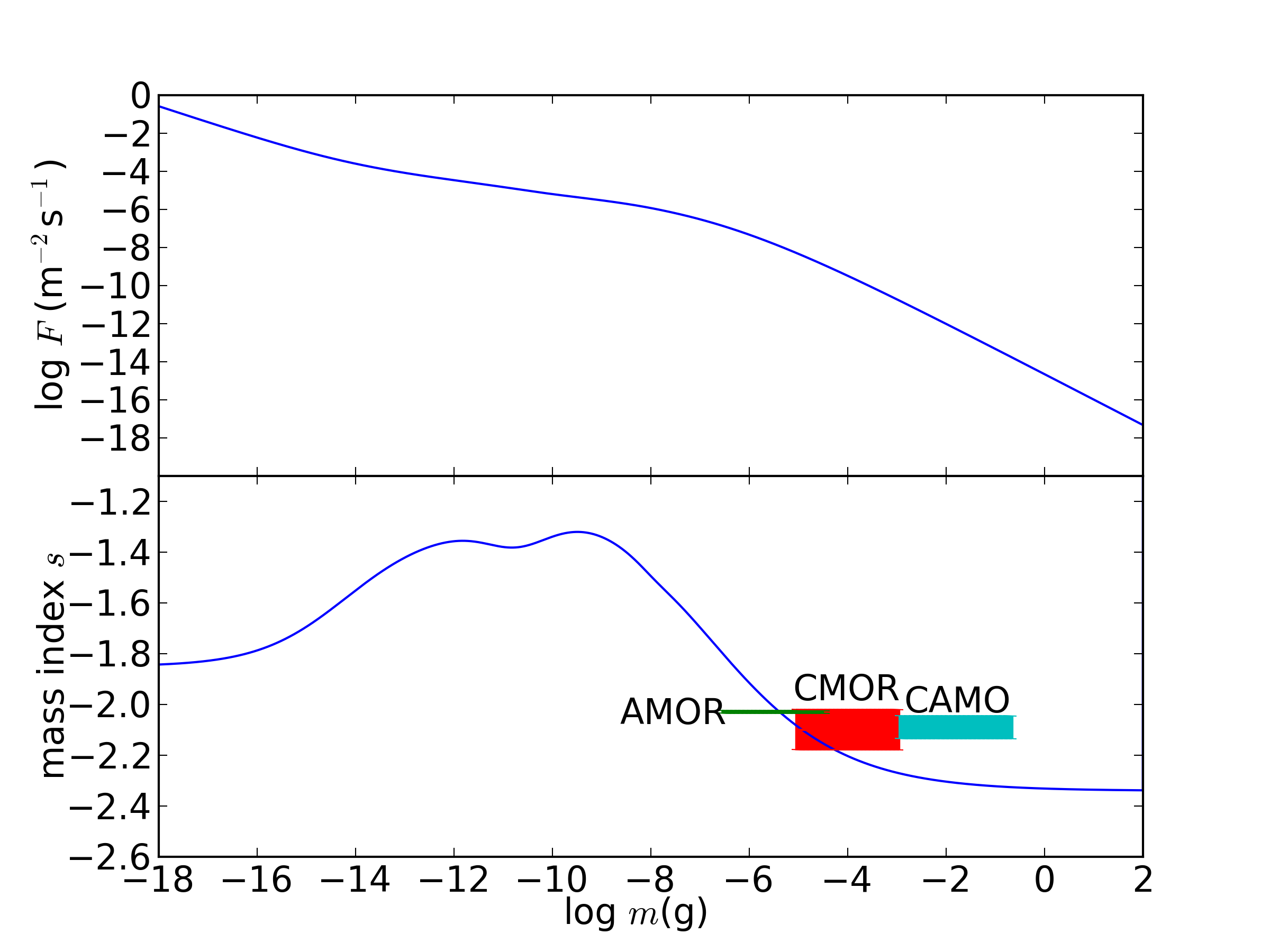}
      \caption{Logarithm of cumulative particle flux at 1 AU distance from the Sun with respect to the logarithm of meteoroid mass adapted from 
      \citet{Grun_etal_1985} (top panel). Mass index $s$ for meteoroids at 1 AU distance from the Sun derived as the first derivative from the 
      curve in the top panel (solid blue line) compared to the mass index $s = 2.03 \pm 0.01$ for meteoroid environment measured by AMOR in 1995--1999 
      \citep{Galligan_Baggaley_2004}, mass index $s = 2.05 \pm 0.08$ for background meteoroid environment from CMOR multi-frequency data, and with mass index $s = 2.09 \pm 0.05$ 
      derived from CAMO observations.}
         \label{FIG_605}
   \end{figure}

The direct comparison of our findings with the Grün curve is shown in Fig. \ref{FIG_605}.  Here the top panel shows the original flux curve of 
\citet{Grun_etal_1985} appropriate to 1 AU from the Sun as a function of the meteoroid mass. The bottom panel of Fig. \ref{FIG_605} shows the 
equivalent mass index $s$ of the Grün curve (blue solid line) compared with our findings for CMOR $s = -2.05$ (green solid line) in the 
range $10^{-4.5}$ to $10^{-3.5}$ g, and with CAMO $s = -2.08$ (red solid line) in the range $10^{-3.6}$ to $10^{-0.6}$g. Our data show a 
systematically shallower mass index compared to \citet{Hawkins_Upton_1958}, as well as the Grun curve at our masses.

\citet{Galligan_Baggaley_2004} examined more than half a million high-quality meteor echoes from the Advanced Meteor Orbit Radar (AMOR) and 
found $s = -2.027 \pm 0.006$. AMOR had a limiting radio magnitude of +14, corresponding to a limiting mass of roughly  $3 \times 10^{-7}$g, 
approximately two orders of magnitude lower than CMOR. Surprisingly, their value is very close to our results. We suggest that this implies 
the turnover to shallower values of the mass index, which only begins near $m \sim 10^{-5}$g in \citet{Grun_etal_1985}, should actually occur 
at roughly an order of magnitude larger mass or that the asymptotic mass index at larger masses should be smaller than the canonical $s=2.34$ 
value. Alternatively the absolute mass scales may be offset.

\section{Conclusions}

We have applied a new fully automated Bayesian-approach based on the MultiNest software package to the  measurement of meteoroid mass indices. 
Application of this technique to CMOR echo amplitude distributions yields a mean overall value for the mass index of $s=-2.05 \pm 0.08$, 
with a yearly variation showing an amplitude of $\pm 0.1$. We note that major apparent “dips” in the mass index throughout the year are 
correlated with a handful of the strongest showers, consistent with the fact that most meteor showers have much lower mass indices than 
the sporadic background \citep{Jenniskens_2006}.

We successfully applied our code on data from two single-station radars performing measurements at 29.85 and 38.15 MHz during 2011--2015. 
After applying several filters for the range and height of the meteor and removing other known biases, we obtained diurnal variations of the 
mass index $s$ for the meteoroid complex seen by CMOR. The mass index undergoes noticeable variations that are beyond uncertainty ranges 
obtained by our algorithm, where several dips in the mass index value were associated with known major meteors showers (Fig. \ref{FIG_601}). 
The average value for the whole meteoroid complex for the 29.85 MHz radar is $s =-2.015 \pm 0.072$, and for the 38.15 MHz radar $s =-2.080 \pm 0.075$. 
Our adopted best fit observed that the radar-determined mass index is $s=-2.05 \pm 0.08$. 

Optical measurements from the influx system of CAMO produced a sporadic-only mass index of $s=-2.08 \pm 0.06$. Applying initial radius biases 
to the CAMO “true” distribution suggests our radar mass index is minimally affected by ITR corrections, with a systematic shift to larger 
values of $s$ of no more than 0.05, so that a final “corrected” mass index is best estimated as being $s=-2.1 \pm 0.08$.

These numbers are within uncertainty ranges comparable with works of \citet{Blaauw_etal_2011} and \citet{Galligan_Baggaley_2004}. Our mass 
index is shallower than that used in \citet{Grun_etal_1985} and we suggest that the turnover to shallower values of the mass index, which only 
begins near $m \sim 10^{-5}$g in \citet{Grun_etal_1985}, should actually occur at roughly an order of magnitude larger mass or that the value 
for $s$ at large mass should be smaller than the 2.34 value commonly quoted.

\begin{acknowledgements}
      The code we developed is based on the MultiNest package that is available (see: https://ccpforge.cse.rl.ac.uk/gf/project/multinest/). 
      Our own code with a simple manual and a sample dataset can be found here: ftp://aquarid.physics.uwo.ca/pub/peter/MassIndexCode/. 
      PGB thanks the Canada Research Chair program. This work is supported in part by the Natural Sciences and Engineering Research Council of 
      Canada and by the NASA Meteoroid Environment Office through NASA co-operative agreement NN15AC94A.
      We would like to thank Althea Moorhead and Laura Lenki\'{c} for helpful comments.
\end{acknowledgements}

%
   \bibliographystyle{aa} 
   \bibliography{BIBTEX} 
%

%
%

\end{document}